# Towards a Cybersecurity Testbed for Agricultural Vehicles and Environments

**Mark Freyhof**
University of Nebraska-Lincoln
mfreyhof2@huskers.unl.edu

**George Grispos**
University of Nebraska-Omaha
ggrispos@unomaha.edu

**Santosh Pitla**
University of Nebraska-Lincoln
spitla2@unl.edu

**Cody Stolle**
University of Nebraska-Lincoln
cstolle2@unl.edu

**ABSTRACT**

In today's modern farm, an increasing number of agricultural systems and vehicles are connected to the Internet. While the benefits of networked agricultural machinery are attractive, this technological shift is also creating an environment that is conducive to cyberattacks. While previous research has focused on general cybersecurity concerns in the farming and agricultural industries, minimal research has focused on techniques for identifying security vulnerabilities within actual agricultural systems that could be exploited by cybercriminals. Hence, this paper presents STAVE – a Security Testbed for Agricultural Vehicles and Environments – as a potential solution to assist with the identification of cybersecurity vulnerabilities within commercially available off-the-shelf components used in certain agricultural systems. This paper reports ongoing research efforts to develop and refine the STAVE testbed, along with describing initial cybersecurity experimentation which aims to identify security vulnerabilities within wireless and Controller Area Network (CAN) Bus agricultural vehicle components.

**INTRODUCTION**

The amalgamation of information technology into agricultural settings introduces a variety of opportunities for the farming community. Agricultural machinery, such as tractors, planters, and harvesting equipment, which once operated as standalone machines, are now being integrated with computer networking technology (Ayaz et al. 2019). However, while the benefits of networked agricultural machinery are attractive (Barreto and Amaral 2018), this technological shift introduces a variety of cybersecurity challenges (Demestichas et al. 2020; Gyles 2010; Yazdinejad et al. 2021). These challenges include a malicious actor gaining control of on-field sensors, and machinery, unauthorized changes to data to deceive a farmer, and the potential contamination of food products.

The reality is that the farming and agricultural industries have become a focus for cybercriminals and nation-state actors. In 2021, The Federal Bureau of Investigation (FBI) released a Private Industry Notification that certain state actors were targeting the food and agriculture industries (The Federal Bureau of Investigation Cyber Division 2021). These concerns became a reality when a targeted ransomware attack forced a fifth of the beef processing plants in the United States to shut down, with one organization paying an $11 million ransom to cybercriminals to avoid any further disruption (Bunge 2021; McCrimmon and Matishak 2021).

As cybercrimes increasingly impact the farming and agriculture communities, both industry and academia are attempting to develop tools and strategies for identifying, evaluating, and mitigating cybersecurity concerns related to networked machinery and farming vehicles. While previous research (Sontowski et al. 2020; Window 2019; Yazdinejad et al. 2021) has focused on the broader cybersecurity challenges associated with agricultural systems connected to the Internet, minimal research focuses on identifying specific security vulnerabilities in agricultural vehicles and machinery, which could be exploited by cybercriminals and nation-state actors.

This paper presents STAVE – a Security Testbed for Agricultural Vehicles and Environments, as a potential solution to assist with the identification of cybersecurity vulnerabilities within specific farming machinery. Moreover, this paper reports ongoing research efforts to develop and refine the STAVE testbed, along with describing initial experimentation to identify potential security vulnerabilities within wireless and Controller Area Network (CAN) Bus agricultural vehicle components. The paper is structured as follows. The next section justifies developing a testbed to investigate cybersecurity vulnerabilities, instead of examining real-world agricultural vehicles or systems. The third section presents STAVE and how the testbed was developed, while the fourth section presents an overview of initial cybersecurity experimentation. The final section concludes the paper and presents ideas for future research.

**PROBLEM DEFINITION**

Agricultural vehicle manufacturers are no longer producing machinery that are isolated systems. For example, in January 2022 John Deere announced that it would begin to mass-produce a self-driving tractor connected to the Internet, which a farmer could also control using a smartphone (Tibken 2022). From a cybersecurity perspective, this increased connectivity and the integration of multiple technologies magnify the available attack surfaces that could be exploited by malicious actors (Grispos 2019a, 2019b). One response is to implement cybersecurity controls and countermeasures, which can reduce or eliminate the potential vulnerability that could exist in a particular agricultural vehicle or machine. However, before these controls and countermeasures can be applied, there is first, a need to identify that a vulnerability exists. These activities are usually considered as part of the security development lifecycle for many software systems (Howard 2004; Lipner 2010), including automotive software systems (Dobaj et al. 2021). Usually, security engineers will study the real software system under development and perform penetration testing (security testing) to identify any cybersecurity weaknesses. However, the very nature of agricultural vehicles and machinery (e.g., their complexity and physical size) means that performing any security testing can be both cumbersome, and impractical (Fowler et al. 2017). Further complicating matters, researchers are also likely to be interested in investigating the latest and most recent agricultural vehicles or machinery available on the market, but these are likely to be very expensive. As a result, researchers will likely investigate older vehicles or machinery instead (Fowler et al. 2017). The wider repercussion is that cybersecurity researchers will propose outdated and incomplete security measures.

**STAVE – A SECURITY TESTBED FOR AGRICULTURAL VEHICLES AND ENVIRONMENTS**

STAVE is proposed as one solution to help identify and investigate cybersecurity vulnerabilities within emerging autonomous agricultural vehicles. As a testbed, STAVE has been developed to mimic an agricultural vehicle called Flex-Ro (Murman 2019; Werner 2016), as shown in Figure 1. This supervised autonomous agricultural vehicle was developed to perform low-draft agricultural operations and is powered by a 57 hp gasoline engine. Flex-Ro was developed using commercially available off-the-shelf components including multiple Danfoss Electronic Control Units (ECUs), which communicate on a CAN bus network operating under the SAE J1939 protocol. Global Positioning System (GPS) allows Flex-Ro to implement precise navigation and operate autonomously. This agricultural vehicle can be operated over a wireless CAN bridge or driven using a laptop computer that can be physically connected to the vehicle. A Farmobile telematics unit is used to transmit machine operation data remotely, back to the user.

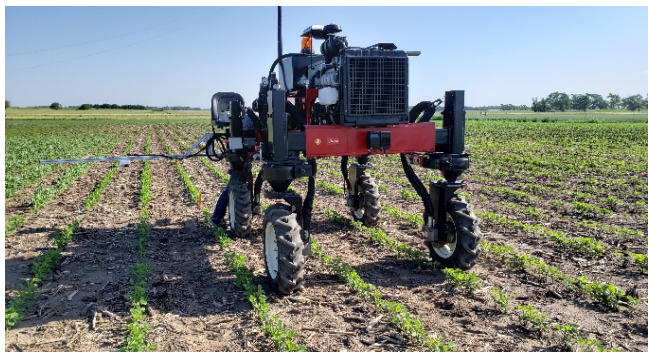

**Figure 1. Flex-Ro in a Field**

Due to Flex-Ro's physical size and complexity, a testbed was developed to allow the identification and investigation of cybersecurity vulnerabilities within the electronic control components used in Flex-Ro. The actual components included in the testbed, along with their functions, are presented below and are numbered in Figure 2 – STAVE Testbed:

(1) Implement ECU (Danfoss MC024-110) – sends output signals to LED lights and motors.

(2) Hydraulics ECU (Danfoss MC024-110) – controls hydraulic pumps to control machine movement.

(3) Engine ECU (Danfoss MC024-110) – sends and receives engine messages.

(4) Power ECU (Danfoss MC012-010) – transmits machine voltage and enables steer motors.

(5) Steering ECU (Danfoss MC012-010) – controls electric steer motor based on inputs.

(6) Wireless CAN Bridge (Magnetek WIC-2402) – creates a wireless connection where CAN messages can be sent and received.

In addition to the above components, a Danfoss DP600TM display ECU and Danfoss JS1000 Joystick (both not shown in Figure 2), were also included in the testbed. A pragmatic decision was made to exclude certain components (e.g., the GPS and Farmobile telematics unit) from the testbed, as initial investigations will focus on potential cybersecurity vulnerabilities across the wireless CAN bridge. Further components will be added to the testbed at a future date. Moreover, the software used to program the testbed ECUs is also identical to the software used to program the Flex-Ro ECUs. To investigate CAN message security, a laptop and a Raspberry Pi were added to the testbed to allow for the monitoring and transmission of CAN bus messages. The absence of a gasoline engine in the testbed results in the need to include LED lights to confirm that the testbed ECUs are indeed transmitting control signals.

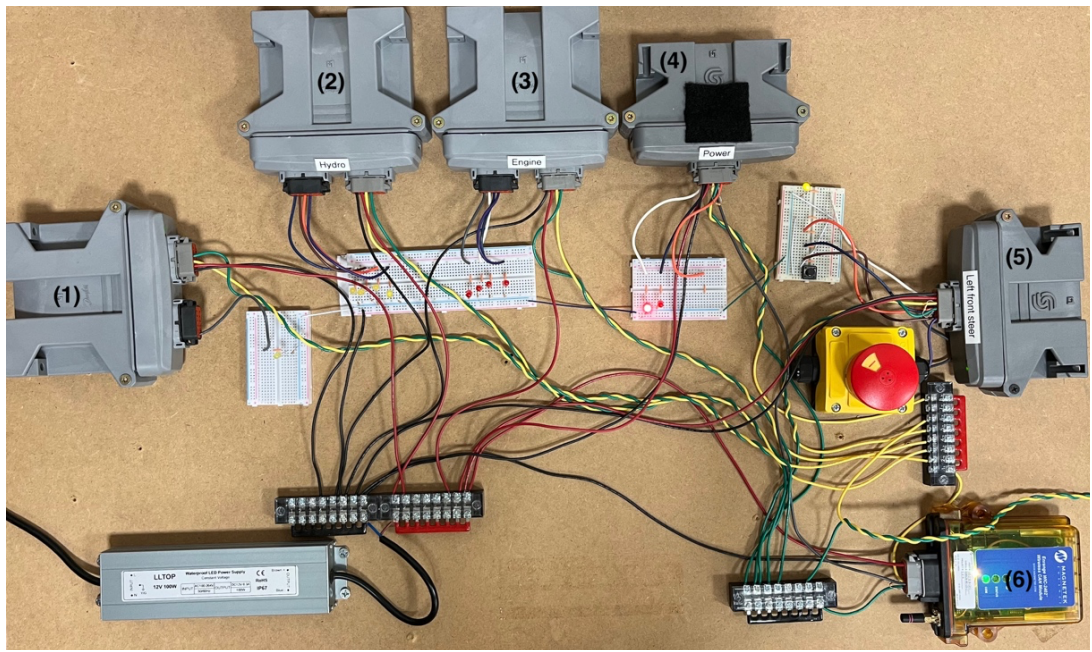

**Figure 2. STAVE Testbed**

**INITIAL EXPERIMENTATION USING STAVE**

Initial experimentation using STAVE focuses on potential security exploitation of the WIC (Wireless CAN) module and performing a replay attack on the CAN bus network. Traditionally, a replay attack occurs when a malicious actor eavesdrops on network communications, intercepts network messages, and then delays or resends these messages to the intended receiver (Harris 2010). From a CAN bus perspective, this involves capturing messages from the CAN bus, modifying, or altering the message, and then retransmitting these modified messages 'back down' the CAN bus network. At a high level, the objective of such an attack is to modify or disrupt the normal operation of the vehicle under attack (Wang and Sawhney 2014). The investigation of this potential exploit is examined in two phases.

The first phase of the experimentation focuses on the exploitation of the actual wireless CAN bridge, i.e., allowing an attacker to obtain access to the CAN bus through the WIC-2402 component. Hence, this part of the experimentation involves the identification of vulnerabilities in this component. This includes the identification and digestion of open-source documentation related to the WIC-2402 to determine what frequency the WIC uses and whether the WIC makes use of frequency hopping. Figure 3 shows wireless signals being captured using a spectrum analyzer, to identify the frequency or 'peaks' used by the signal originating from the WIC-2402 component.

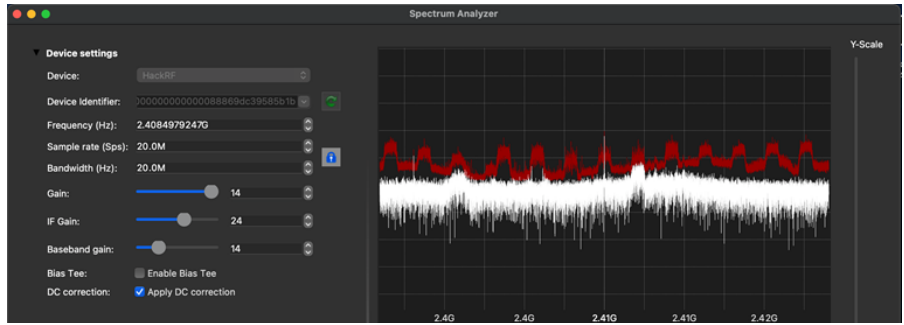

**Figure 3. Wireless Signals Captured Using Spectrum Analyzer**

After identifying the frequency range used to send and receive wireless packets, a HackRF device (Great Scott Gadgets n.d.) and Universal Radio Hacker (URH) software (Pohl and Noack 2018) can be used to 'sniff' wireless packets transmitted from the STAVE testbed's WIC component. However, the popularity of certain wireless frequencies (e.g., 2.4GHz) complicates attempts to isolate individual wireless packets from a specific device. Hence, the use of a Faraday box (shown in Figure 4) is needed to isolate the wireless packets from the WIC-2402, with those originating from other wireless devices that also use a similar frequency range.

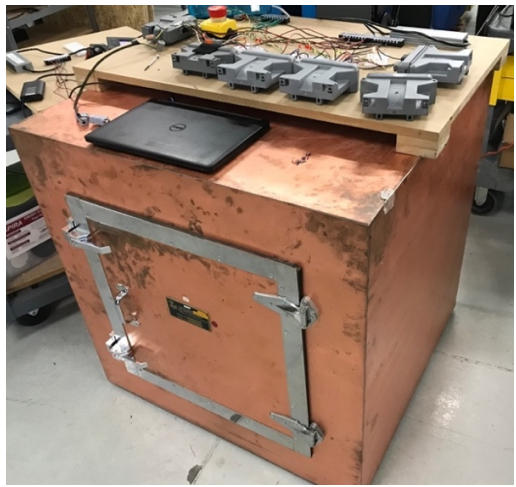

**Figure 4. Faraday Box used in STAVE Testbed**

The second phase of the experimentation involves the capture and examination of CAN bus network packets. A CAN bus shield is used to connect a Raspberry Pi executing Wireshark to STAVE, to capture the CAN bus packets in a pre-test/post-test manner (Campbell and Stanley 2015). These CAN bus packets are collected when the joystick and the steering ECU are idle, and then recaptured when the joystick is being used, and the steering ECU is active. The difference between the idle packets (pre-test) and the packets when the joystick is being used (post-test), allows the identification of information in the CAN bus packets about the joystick manipulation and how this information is transmitted on the CAN bus network. Once this information can be identified, it can be modified to manipulate the steering of the potential agricultural vehicle, i.e., if the joystick is moved to the left, is it possible to instruct the ECU to direct the agricultural vehicle to the right.

The wireless packets, which were captured in the first phase of the experiment, can then be decoded using the URH software and a hex editor. The purpose of this decoding is to locate and identify CAN bus information from the second phase. Finally, using this information, a modified CAN bus network packet can be retransmitted in a replay attack manner back to the WIC-2402, which would transmit the information to the CAN bus network and related ECU.

## CONCLUSIONS AND FUTURE WORK

While previous research has focused on the broader cybersecurity challenges related to the farming and agricultural communities, minimal research has focused on identifying security vulnerabilities in agricultural vehicles and machinery. Hence, this paper presents the initial results of an ongoing research effort to develop a cybersecurity testbed called STAVE, which will allow for the identification of cybersecurity vulnerabilities within specific farming machinery. Hopefully, the proposed research and the STAVE testbed will provide a foundation for future research endeavors in this domain. However, there is much to be done. Future research will focus on improving and refining the STAVE testbed, through additional components and the verification of vulnerabilities identified using the testbed against the real-world Flex-Ro vehicle. Future work will also examine the cybersecurity vulnerabilities associated with cybercriminals exploiting one agricultural vehicle and, potentially, using that vehicle to cause further cyber or physical damage to an individual's farm or agricultural environment. Finally, future research will also focus on identifying security controls to mitigate the vulnerabilities identified using the STAVE testbed.

*Freyhof, et al.* *Towards a Cybersecurity Testbed for Agricultural Vehicles and Environments*
22. Werner, J. P. (2016). "Flex-Ro: Design, Implementation, and Control of Subassemblies for an Agricultural Robotic Platform." University of Nebraska-Lincoln.
23. Window, M. (2019). "Security in Precision Agriculture: Vulnerabilities and Risks of Agricultural Systems." Luleå University of Technology.
24. Yazdinejad, A., Zolfaghari, B., Azmoodeh, A., Dehghantanha, A., Karimipour, H., Fraser, E., Green, A. G., Russell, C., and Duncan, E. (2021). "A Review on Security of Smart Farming and Precision Agriculture: Security Aspects, Attacks, Threats and Countermeasures," *Applied Sciences* (11:16), p. 7518.
*Proceedings of the Seventeenth Midwest Association for Information Systems Conference, Omaha, Nebraska, May 16-17, 2022*     *6*